\documentclass[proof]{WileyASNA-v1}

\articletype{Proceedings}%

\received{20 December 2022}
\revised{06 January 2023}
\accepted{06 January 2023}

\begin{document}

\title{Equations of State for Dense Matter and Astrophysical Constraints}

\author[1]{Rafael Bán Jacobsen}
\author[2]{Verônica Dexheimer}
\author[1]{Ricardo Luciano Sonego Farias}

\authormark{V. Dexheimer, R.B. Jacobsen, R.L.S. Farias}

\address[1]{\orgname{Universidade Federal de Santa Maria (UFSM)}, \orgaddress{\state{Santa Maria}, \country{Brazil}}} 

\address[2]{Department of Physics, \orgname{Kent State University}, \orgaddress{\state{Kent, OH 44243}, \country{USA}}} 

\abstract{ABSTRACT: This conference proceeding presents an overview of the modern approaches in the study of baryonic matter at high densities, focusing on the use of online repositories such as CompOSE and MUSES  for the calculation of neutron star properties. In this context, relevant astrophysical constraints for the equations of state (mass-radius relation, speed of sound, tidal deformability) are discussed.}


\maketitle

\section{General Aspects of the Equation of State for Dense Matter}

The study of the properties of compressed baryonic matter, or, more specifically, strongly interacting matter at high densities, is a mostly relevant topic for current research in Physics, with implications both in the microscopic and in the large scale realms of nature. In the first domain, heavy-ion collision experiments, such as those carried out by the Relativistic Heavy Ion Collider (RHIC) at Brookhaven National Laboratory and the Large Hadron Collider (LHC) at CERN, provide numerous data on the behavior of baryonic matter at extreme conditions of density and temperature. Additionally, in the second domain, astronomic observations of neutron stars, from both orbiting and ground based observatories, spanning the electromagnetic spectrum from $\gamma$-rays to radio wavelengths and now also including gravitational waves, can unveil significant properties of baryonic matter at high densities, since neutron stars contain compressed baryonic matter in their centers. These remnants of massive stars after core-collapse supernova explosions are typically about 12 kilometers across and may contain up to 2 solar masses ($2 {M}_{\odot}$), implying core densities as high as 10 times nuclear saturation density ($\sim 10^{15} g/cm^3$). 

In both cases, linking data to theoretical description of baryonic matter depends on the equation of state (EoS) adopted. In a broad sense, an EoS is a thermodynamic equation relating state 
variables (and usually including the pressure). In the specific field of nuclear astrophysics, it is also expected that an EoS provides a full thermodynamic list of variables (e.g., chemical potentials, entropy per baryon), particle composition of the system (the proportion of the different types of leptons, nucleons, and hyperons), microscopic information (e.g., effective masses and  pairing gaps) and stellar properties (e.g., maximum mass and radius, tidal deformability). 

EoS input tables for astrophysical simulations usually includes baryon number density ($n_B$), charge fraction ($Y_Q$), and temperature ($T$) as independent variables. A 1-dimensional EoS table depends only on the parameter $n_B$ and may describe cold isospin-symmetric matter ($T=0$ and $Y_Q=0.5$), cold neutron matter ($T=0$ and $Y_Q=0.0$), or cold $\beta$-equilibrated matter ($T=0$ and $Y_Q$ determined by the conditions of $\beta$-equilibrium and charge neutrality). A 2-dimensional EoS table depends on two of the three aforementioned independent variables and may describe, for example, dense matter at zero temperature (varying $n_B$ and $Y_Q$ with $T=0$), symmetric matter (varying $n_B$ and $T$ with $Y_Q=0.5$), neutron matter (varying $n_B$ and $T$ with $Y_Q=0$), and $\beta$-equilibrated matter (varying $n_B$ and $T$, and calculating $Y_Q$ according to $\beta$-equilibrium and charge neutrality). Nonetheless, a 3-dimensional EoS table depends on all three free parameters and serves for general purposes. Namely, a 3-dimensional EoS table is required for supernova and mergers simulations as long as, differently from neutron stars, the matter in proto-neutron stars and in hypermassive stars is  hot and not $\beta$-equilibrated.

\begin{figure}[t!]
\centering
\includegraphics[width=68mm]{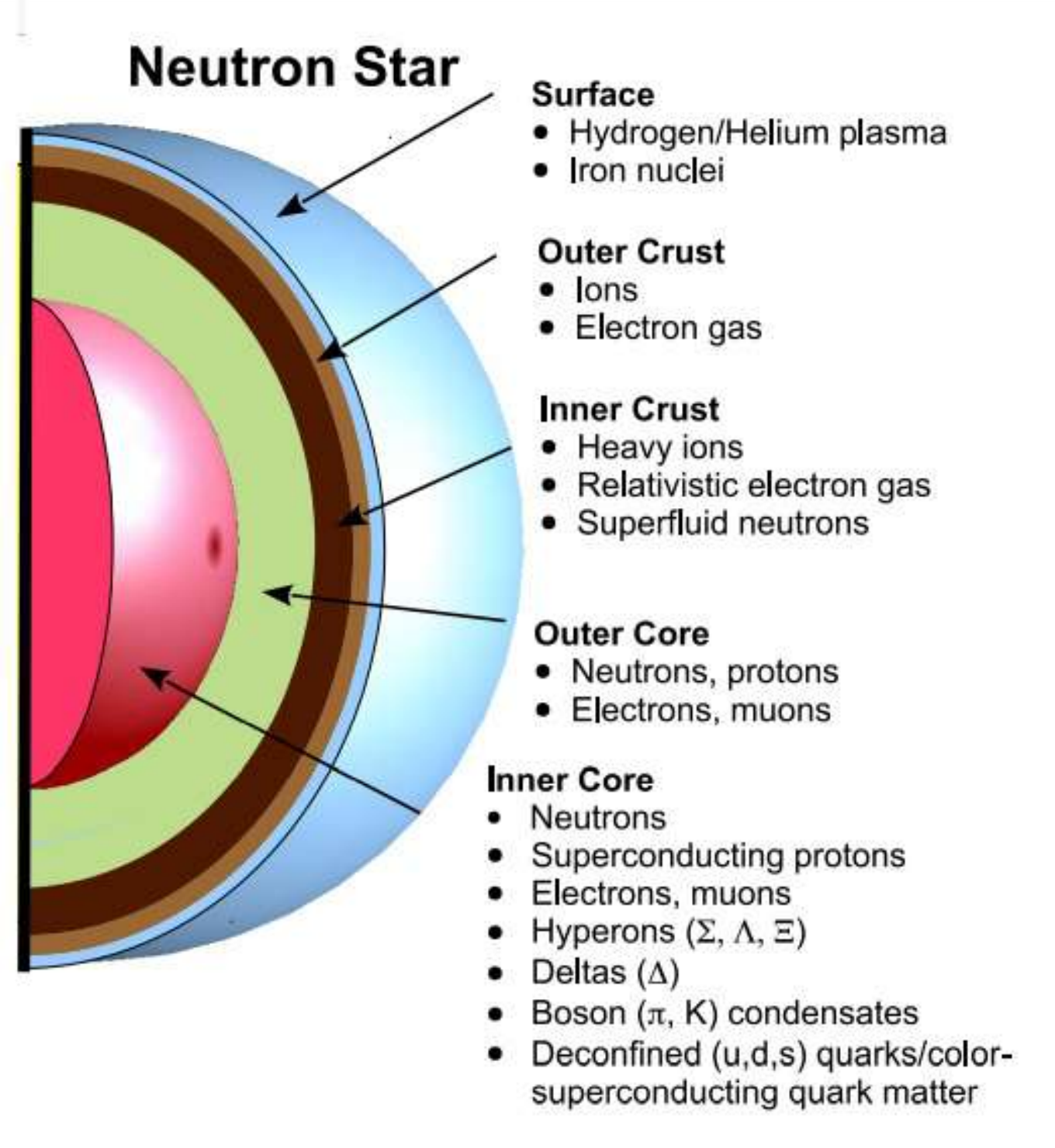} 
\caption{Schematic structure of a neutron star. Figure modified from~\cite{Weber2014}.} \label{FIG1}
\end{figure}

A complete EoS for neutron stars is expected to describe a system with nuclei in the lower density regime, evolving to bulk hadronic matter (nucleons, hyperons, deconfined quarks) at higher densities. Inside neutron stars, this corresponds to the crust and core, respectively (see Fig. \ref{FIG1}).

An EoS for dense and hot matter must be based on a quantum relativistic description, because this framework ensures respect to causality, as long as vector interactions are not too strong. A realistic dense and hot EoS must also obey a series of nuclear and quantum chromodynamics (QCD) constraints:
\begin{itemize}
    \item To reproduce chiral symmetry restoration, as demanded by QCD at large densities and temperatures (with a correspondent decrease in the overall baryonic masses);
    \item To reproduce lattice QCD results at finite temperature (which are provided at any isospin and strangeness, but are restricted to low density relative to the temperature);
    \item To be in agreement with the (nearly) isospin-symmetric and zero net strangeness heavy-ion collision physics at finite temperature;
    \item To reproduce perturbative QCD results in the relevant regime.
    \item To reproduce standard zero-temperature isospin-symmetric nuclear physics results around saturation density. 
\end{itemize}

\section{Modern Sources for Equations of State}

In order to face the challenge of finding an adequate EoS for dense matter in this variety of phenomena, online repositories of equations of state have been built in recent years. CompOSE and MUSES are among these modern sources for 1-, 2-, and 3-dimensional EoS tables.

\subsection{COMPOSE}

CompOSE (CompStar Online Supernovae Equations of State)\footnote{https://compose.obspm.fr} is the largest repository of this kind, offering almost 300 equations of state, divided in families (cold neutron star EoS, cold matter EoS, neutron matter EoS, general purpose EoS, and neutron star crust EoS) and their subgroups (models with hyperons and delta resonances, hybrid quark-hadron models, models with hyperons, models with kaon condensate, nucleonic models, and quark models). The repository also provides a software to interpolate data, calculate additional quantities, and graph EoS dependencies. Data tables, associated software and the manual, can be freely downloaded, cf. \cite{Dexheimer:2022qhn,CompOSECoreTeam:2022ddl}.

\begin{figure}[htb]
\centering
\includegraphics[width=80mm,height=70mm]{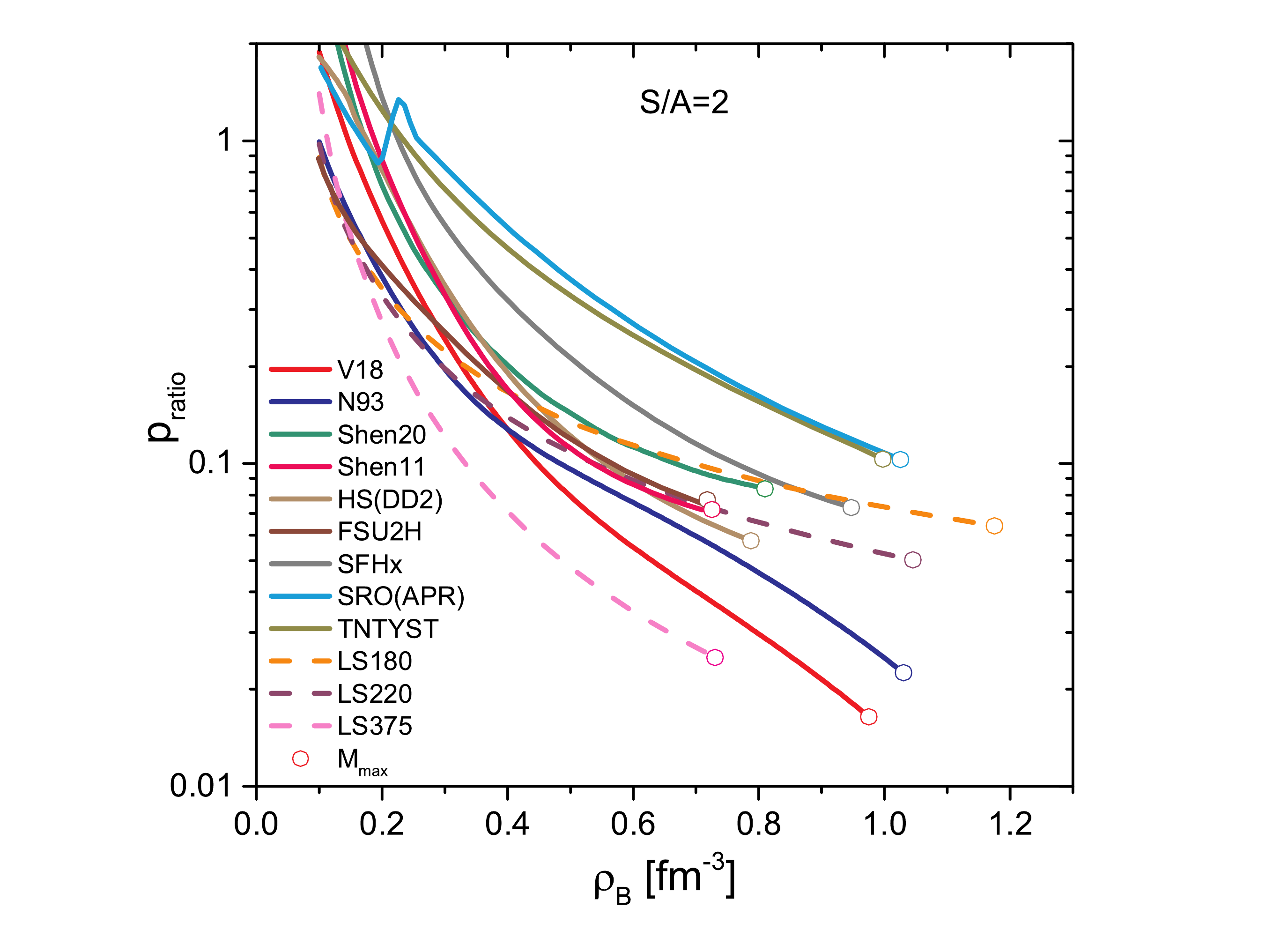}
\includegraphics[width=75mm,height=72mm]{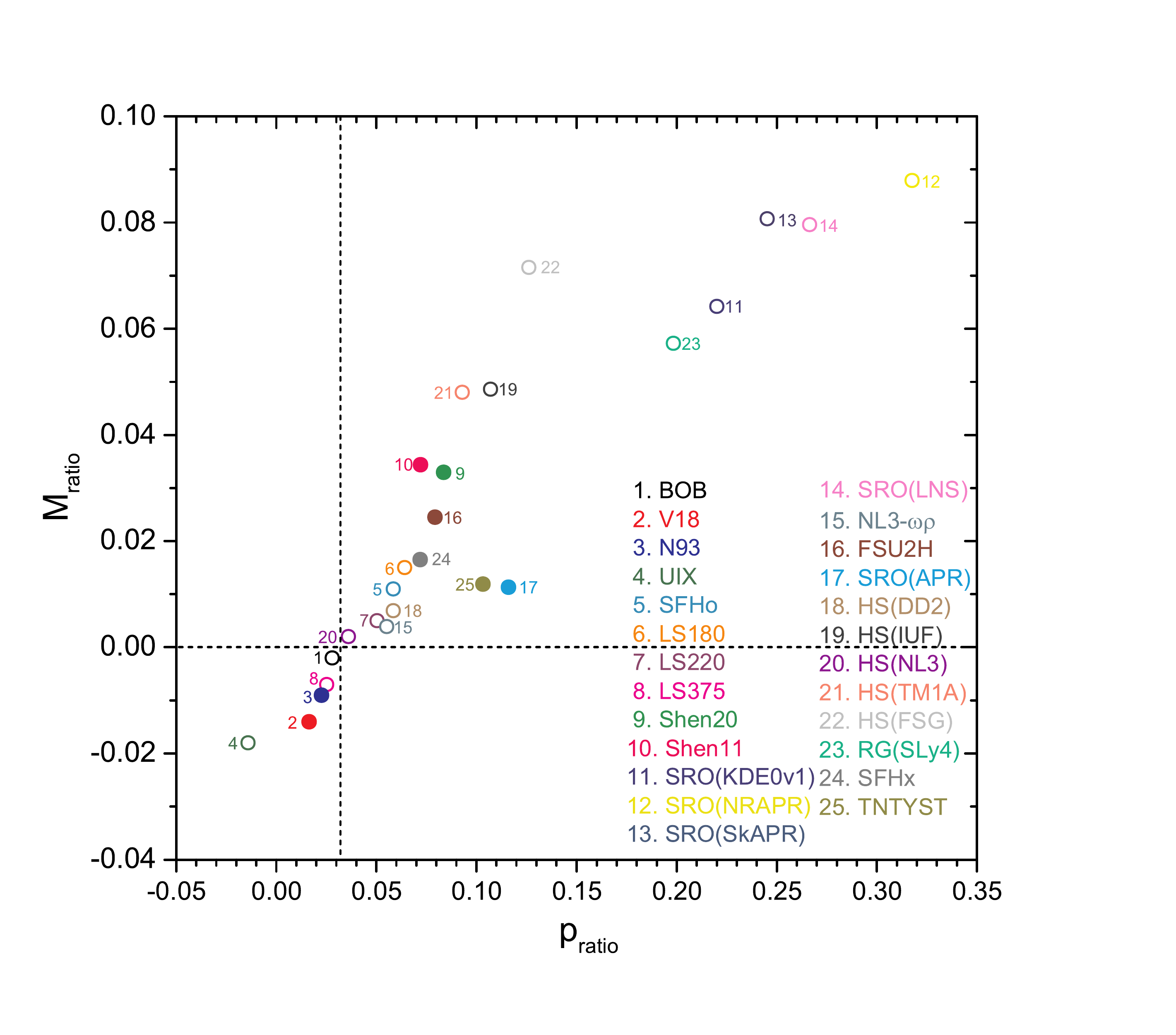}
\caption{Upper panel: Ratio of thermal pressure as a function of density for several equations of state. Lower panel: Relative change of the maximum gravitational mass as a function of the pressure ratio at the center of the star. Figures modified from Fig.6 and Fig.8 in~\citep{Wei_2021}.} \label{FIG2}
\end{figure}

Paradigmatic examples of the usefulness of such a database can be found in studies that carry out comparisons of the predictions made by different models for the same physical system. For instance, a set of microscopic, covariant density-functional, and non-relativistic Skyrme-type equations of state, obtained from CompOSE, has been employed to study the structure of purely nucleonic $\beta$-equilibrated neutron stars at finite temperature~\citep{Wei_2021}. Considering the agreement with presently available astrophysical observational constraints, this study showed that the magnitude of thermal effects depends on the nucleon effective mass as well as on the stiffness of the cold equation of state. Regarding the equations of state themselves, an appropriate quantity to analyze in this context is the relative thermal pressure, defined as ${p}_{ratio} \! =  {p}_{th}/{p}_{0} = \left[p(\rho_B, x_T, T)-p(\rho_B, x_0, 0)\right]/p(\rho_B, x_0, 0)$,
where $\rho_B$ is the baryonic density, $T$ is temperature and $x_{0}$ and $x_{T}$ are the respective proton fractions of cold and hot matter. The ratio of thermal pressure as a function of density is shown in the upper panel of Fig. \ref{FIG2}for the different equations of state studied. 
Moreover, in order to appreciate the astrophysical implications of these equations of state, the relative change of the maximum gravitational neutron-star mass, defined as
${M}_{ratio} \! = \left(M^{hot}_{max}-M^{cold}_{max}\right)/M^{cold}$,
can be plotted as a function of the thermal pressure ratio. The result is shown in the lower panel of Fig. \ref{FIG2}.

\subsection{MUSES}

MUSES (Modular Unified Solver of the Equation of State)\footnote{https://muses.physics.illinois.edu/} is a large collaboration project that is developing a new cyberinfrastructure to provide novel tools to answer critical interdisciplinary questions in nuclear physics, gravitational wave astrophysics and heavy-ion physics. The MUSES collaboration consists of many researchers and technical professionals across dozens of institutions spread across the globe who are building and using a collaborative platform which is \textit{modular} because, while at low baryonic chemical potential the EoS is known from first principles, at high there will be different models for the user to choose; besides, it is \textit{unified} in as much as different modules will be merged together to ensure maximal coverage of the phase diagram. Building up MUSES, physicists and computer scientists will work together to develop the software that generates equations of state over large ranges of temperature and chemical potentials to cover the whole QCD phase diagram. The group of users is composed by interested scientists from different communities, who provide input to the future open-source cyberinfrastructure.

\section{Astrophysical Constraints}

Any consistent EoS has to pass the test posed by the astrophysical constraints related to neutron stars, the most fundamental being the mass-radius relation for these compact objects. Nonetheless, many relevant features cannot be appreciated on such a basis; for example, the possible existence of different exotic matter associated with different phase transitions inside a neutron star can easily be seen in the speed of sound ($c_S$) behavior but not necessarily in the mass-radius relation. As a matter of fact, $2 M_{\odot}$ stars demand a stiff EoS (with ${c_S}^{2} \longrightarrow 1$ in natural units) at intermediate densities; on the other hand, ${c_S}^{2} \longrightarrow 1/3$ from below at asymptotically large densities because of the conformal limit of massless free quarks. Thus, a non-monotonic behavior is expected for $c_S$, implying the occurrence of \textit{bumps} related to the softening of the EoS due to new degrees of freedom, cf.\cite{Bedaque:2014sqa}. 

\begin{figure}
\centering
\includegraphics[width=42.8mm]{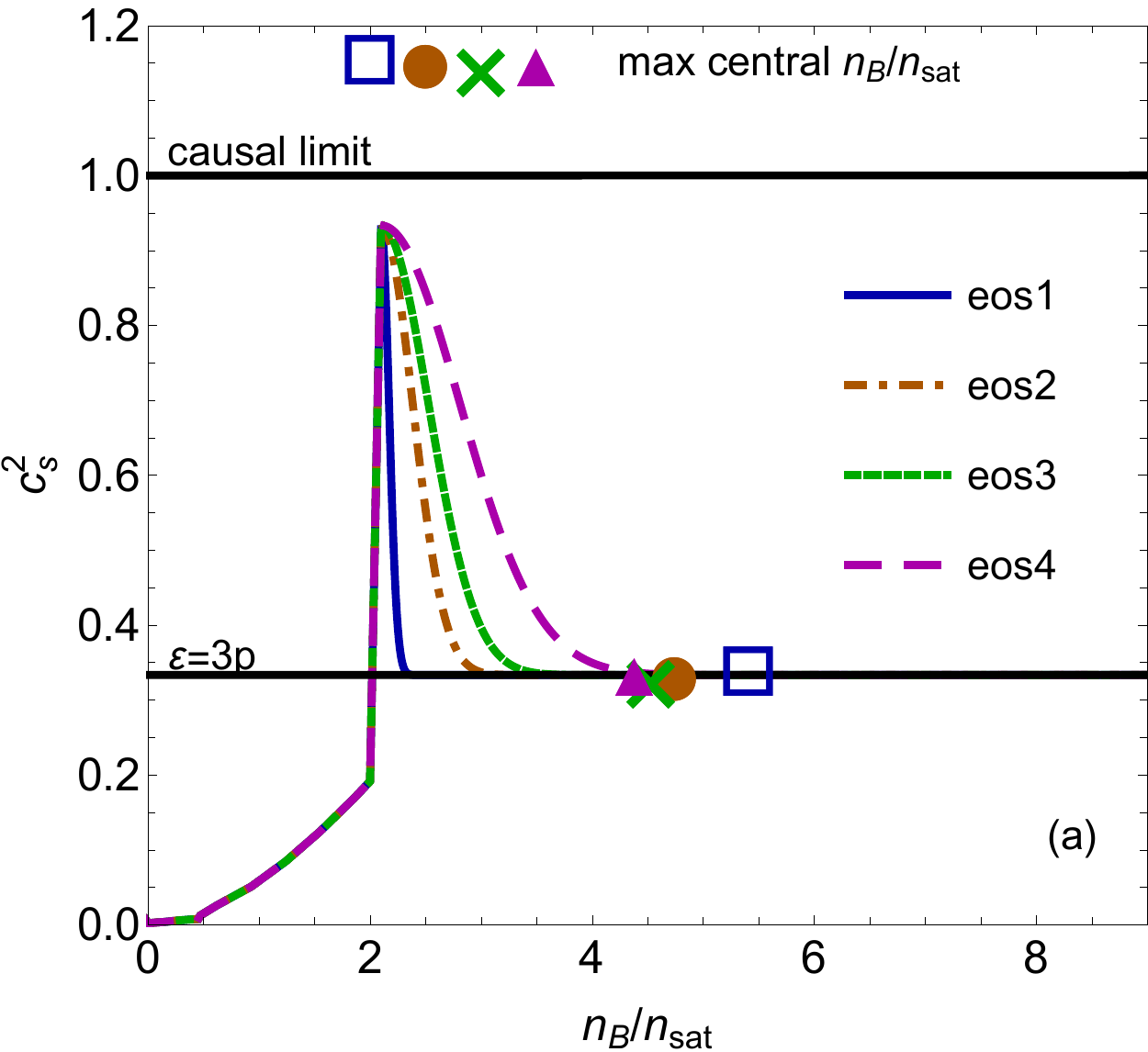}
\includegraphics[width=43.6mm]{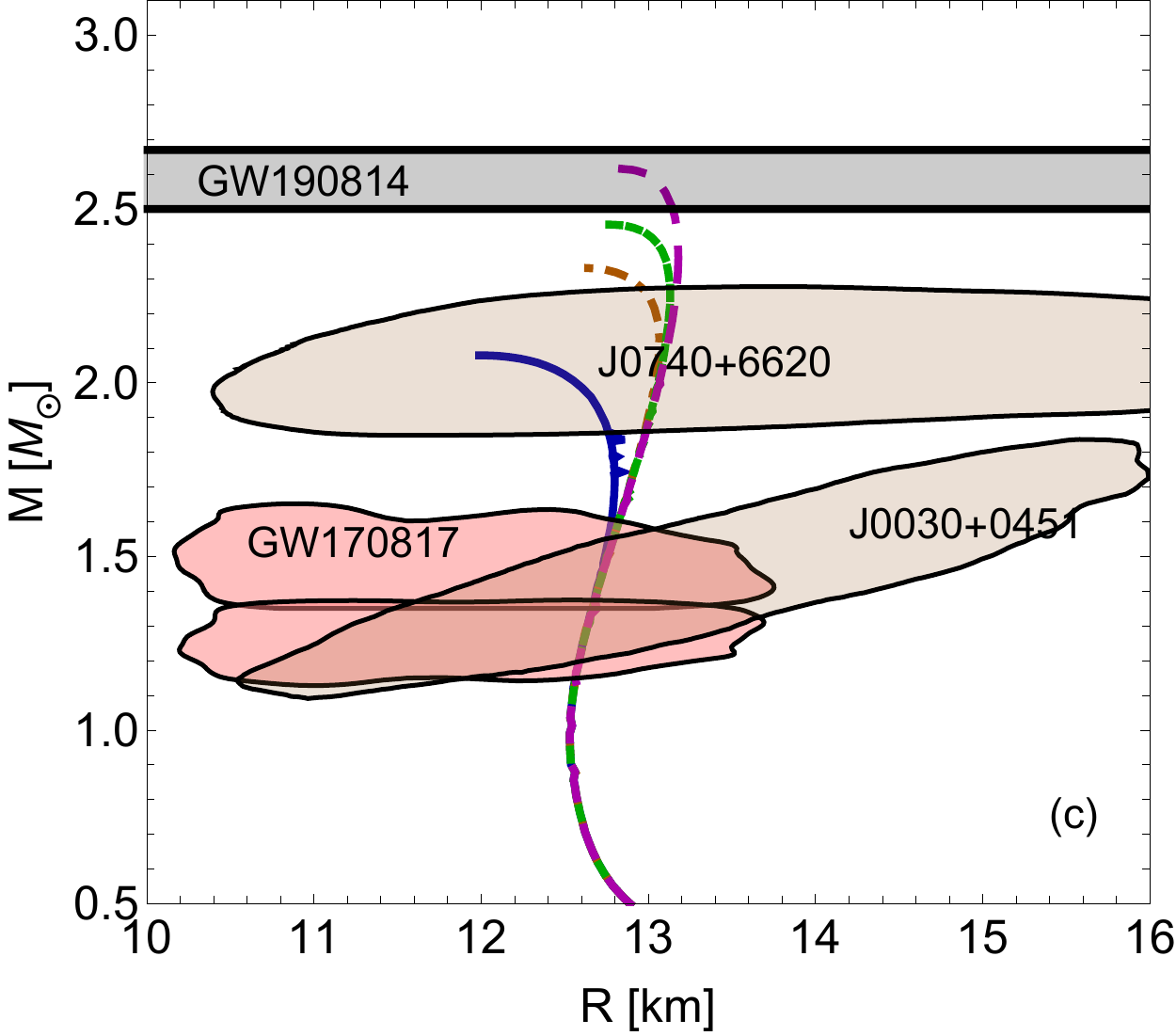}
\includegraphics[width=42.8mm]{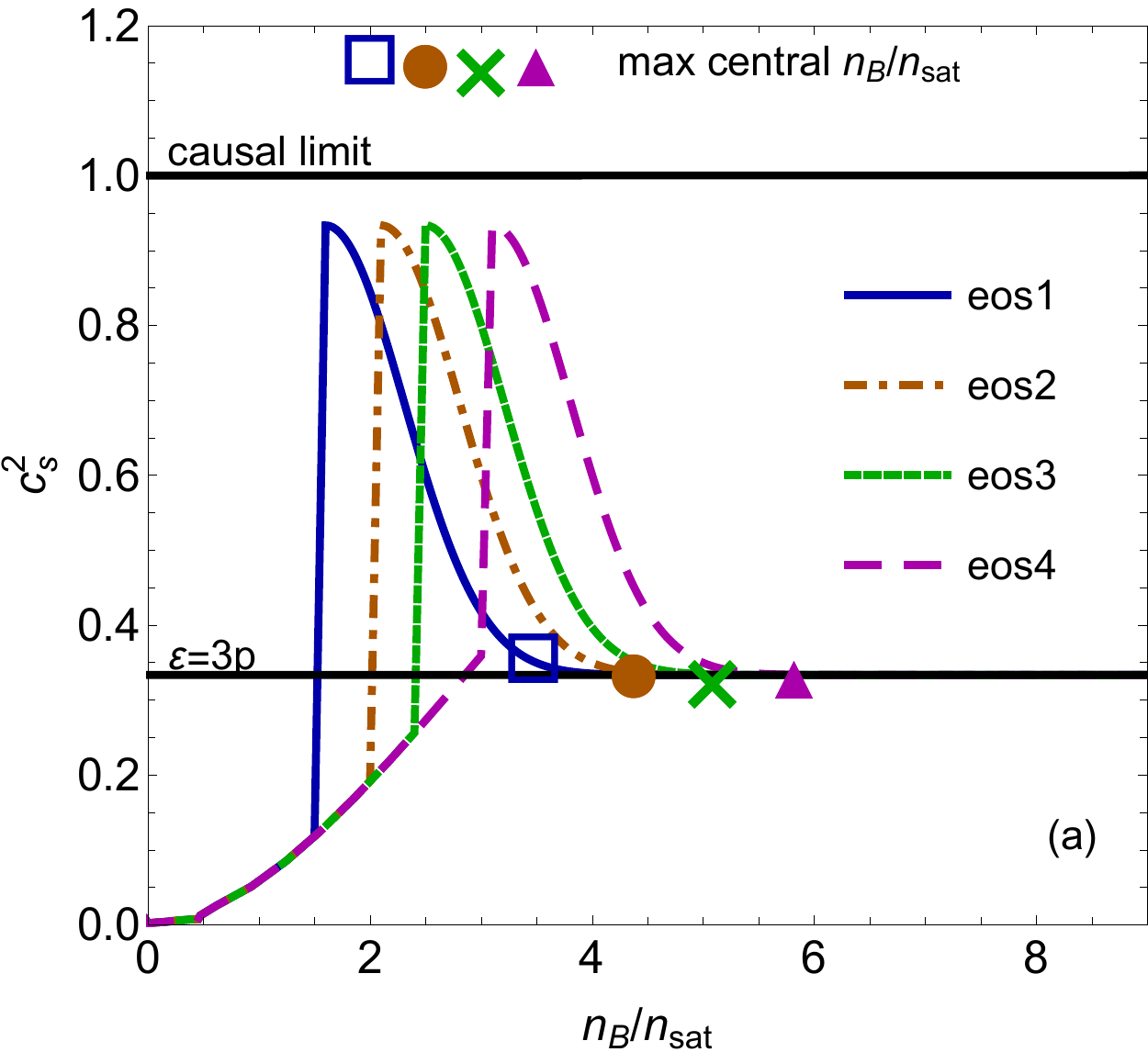}
\includegraphics[width=43.6mm]{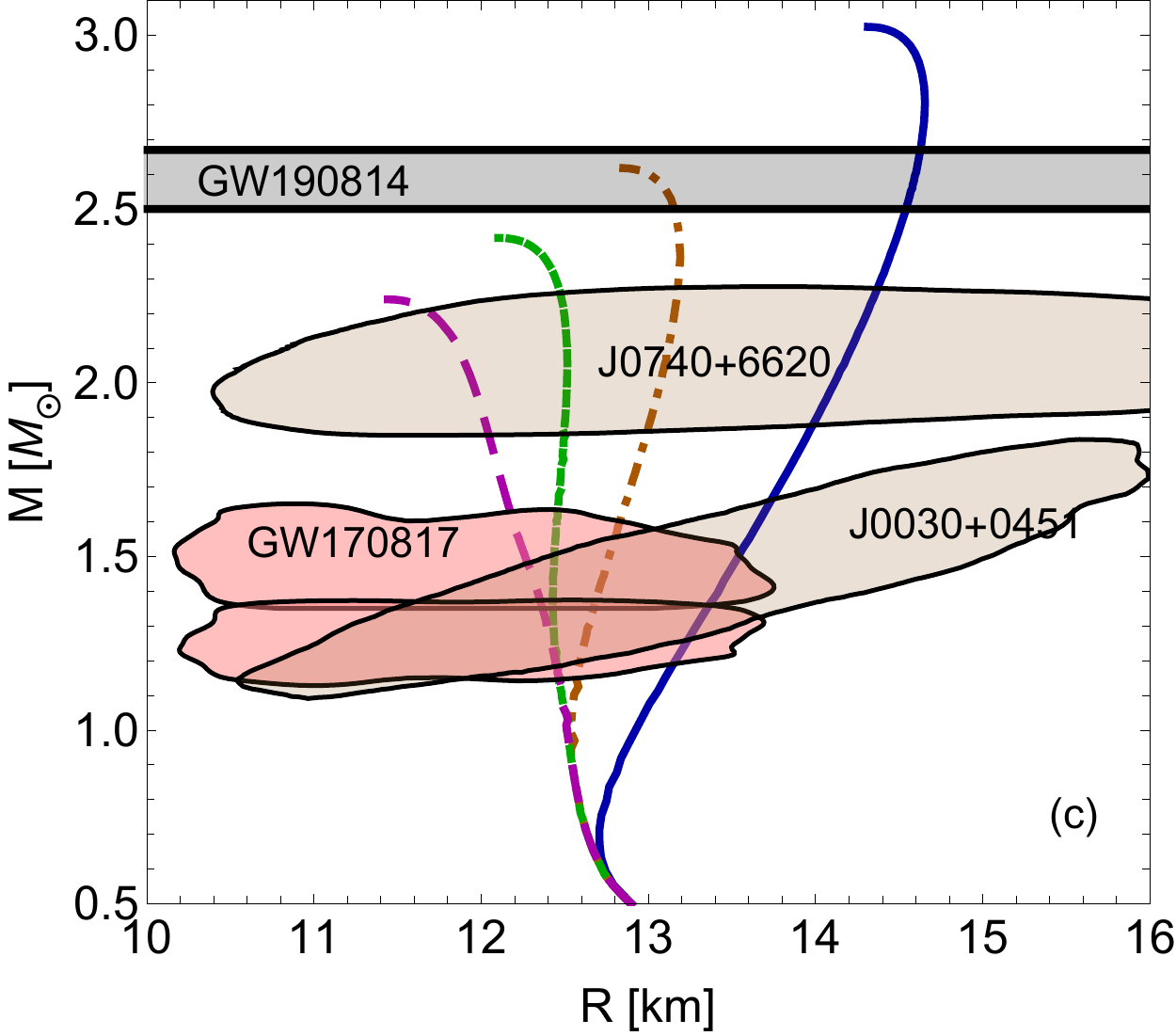}
\caption{Upper panels: Speed of sound (left) and mass-radius diagram (right) for a subfamily of equations of state with peaks in the speed of sound of different widths at the same location. Lower panels: Speed of sound (left) and mass-radius diagram (right) for a subfamily of equations of state with peaks in the speed of sound of the same width at different locations. Modified from Fig.4 in~\cite{Tan_2022}.} \label{FIG4}
\end{figure}

Figure \ref{FIG4}, adapted from~\cite{Tan_2022}, show how bumps (that also appear in realistic microscopic models) can be produced under a controlled $c_s$ parametrization, allowing a correlation between the density at which the bump appears and curves in the neutron star mass-radius diagram. Thus, this more systematic parametric form for the speed of sound can help to determine neutron-star composition; besides, maximum stellar mass and radius can determine width, density, and height of the bumps. The non-smooth structure of the speed of $c_S$ related to phase transitions in dense matter makes feasible the constitution of ultra-heavy neutron stars (with masses larger than $2.5 M_\odot$). These stars pass all observational and theoretical constraints, including those imposed by recent LIGO/Virgo gravitational-wave observations and NICER X-ray observations. 

\begin{figure}[htb]
\centering
\includegraphics[width=42.8mm,trim={0 .65cm 0 0},clip]{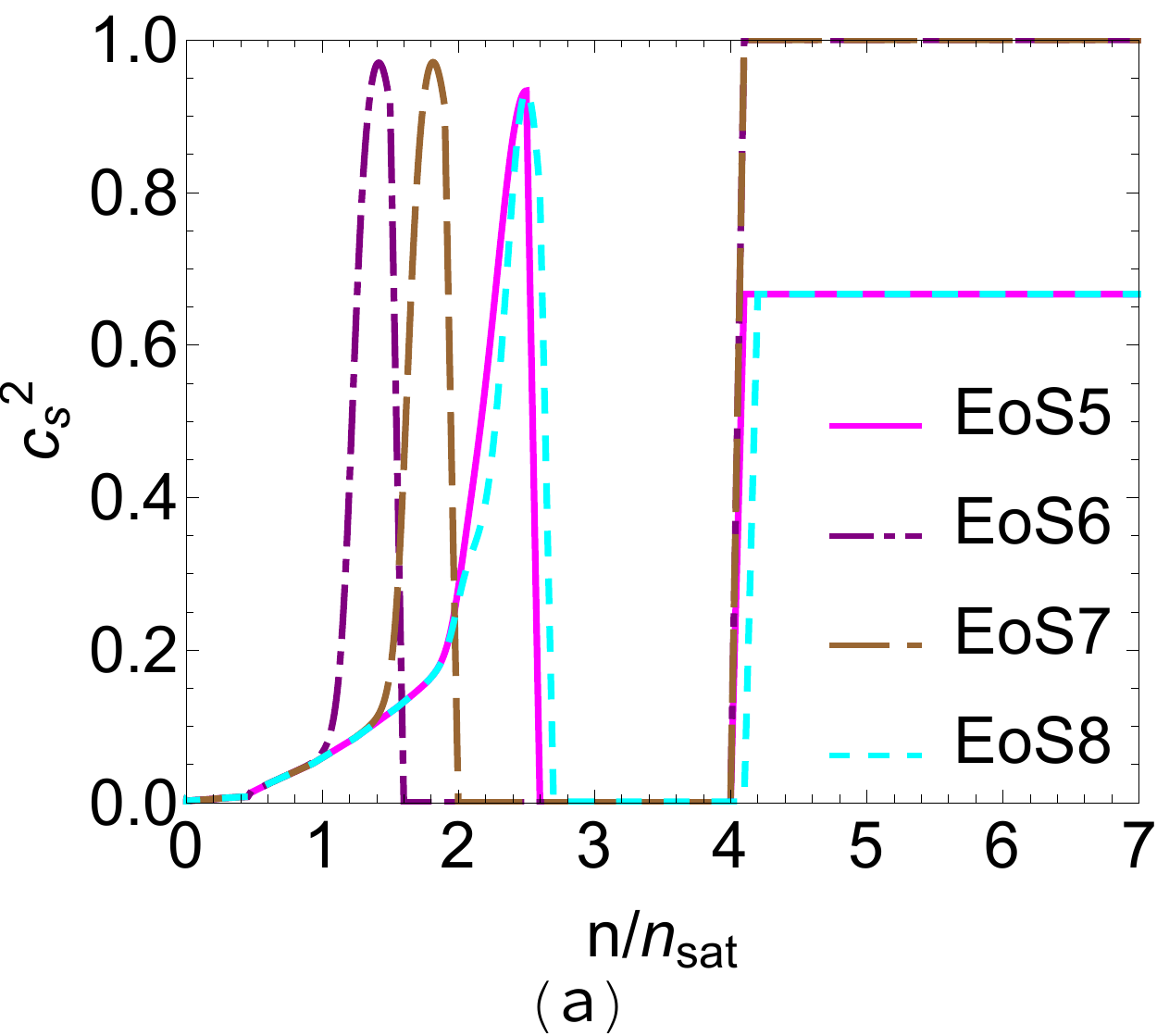}
\includegraphics[width=43.5mm,trim={0 .65cm 0 0},clip]{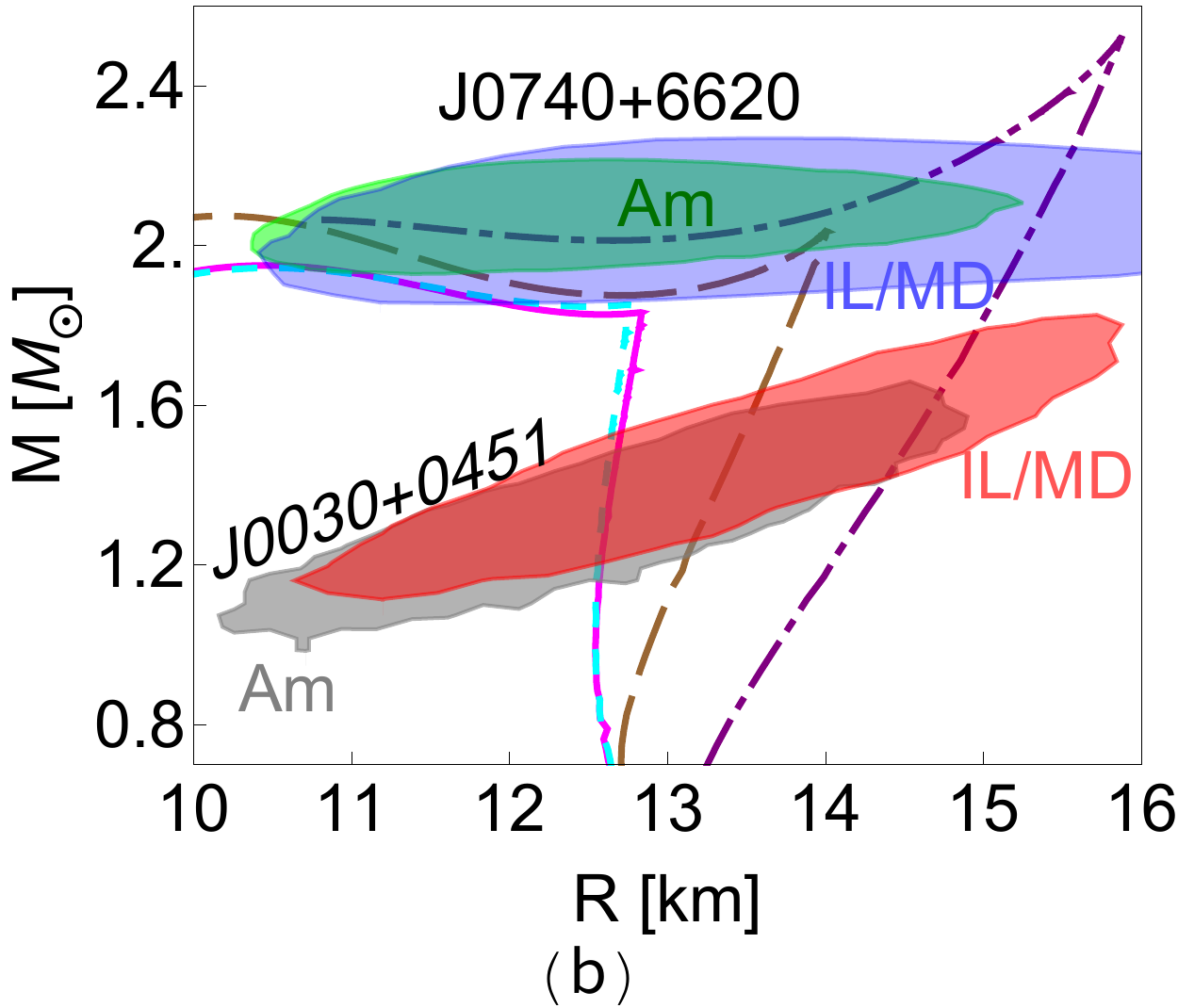}
\includegraphics[width=43.8mm,trim={0 .65cm 0 0},clip]{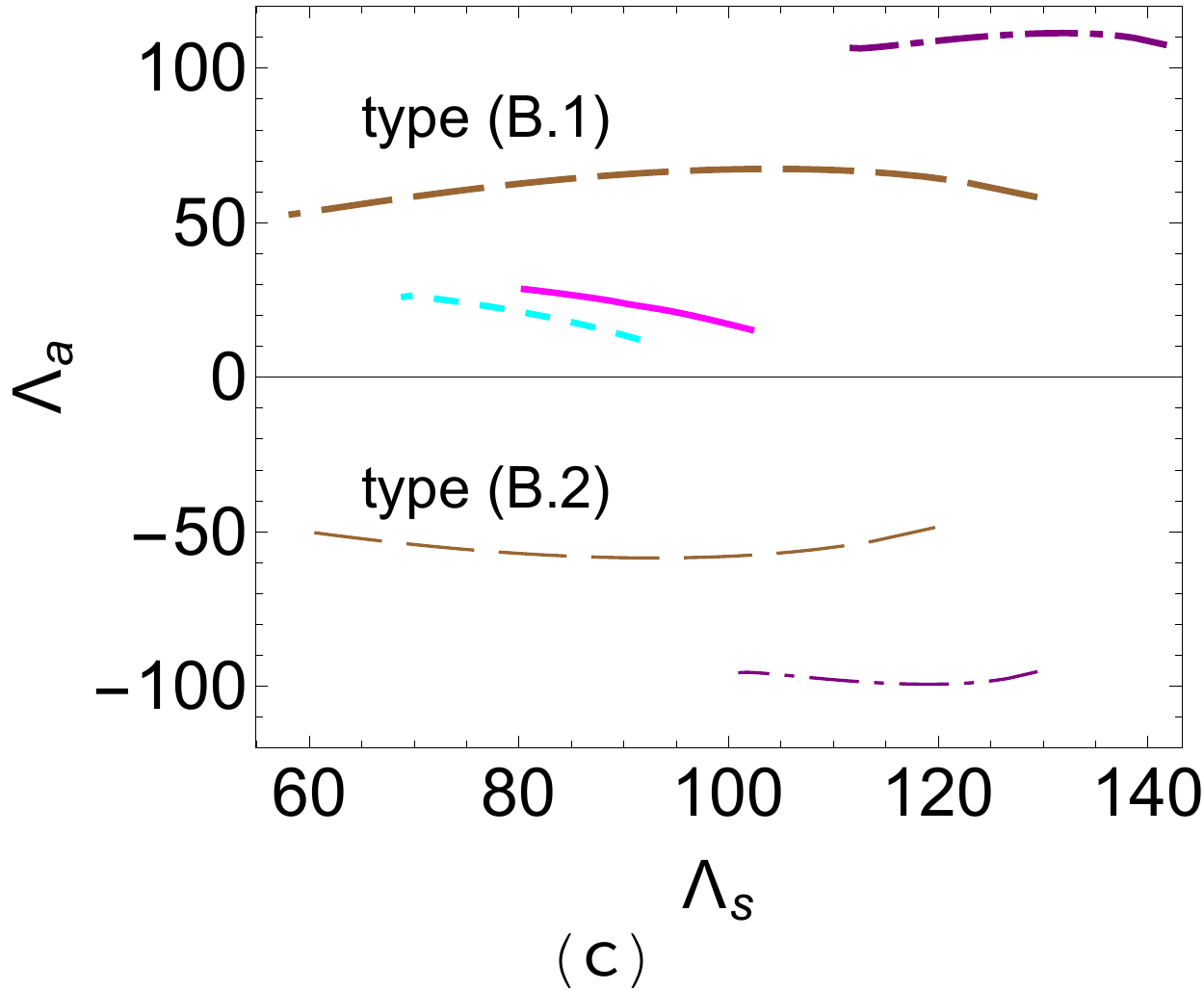}
\includegraphics[width=42.6mm,trim={0 .65cm 0 0},clip]{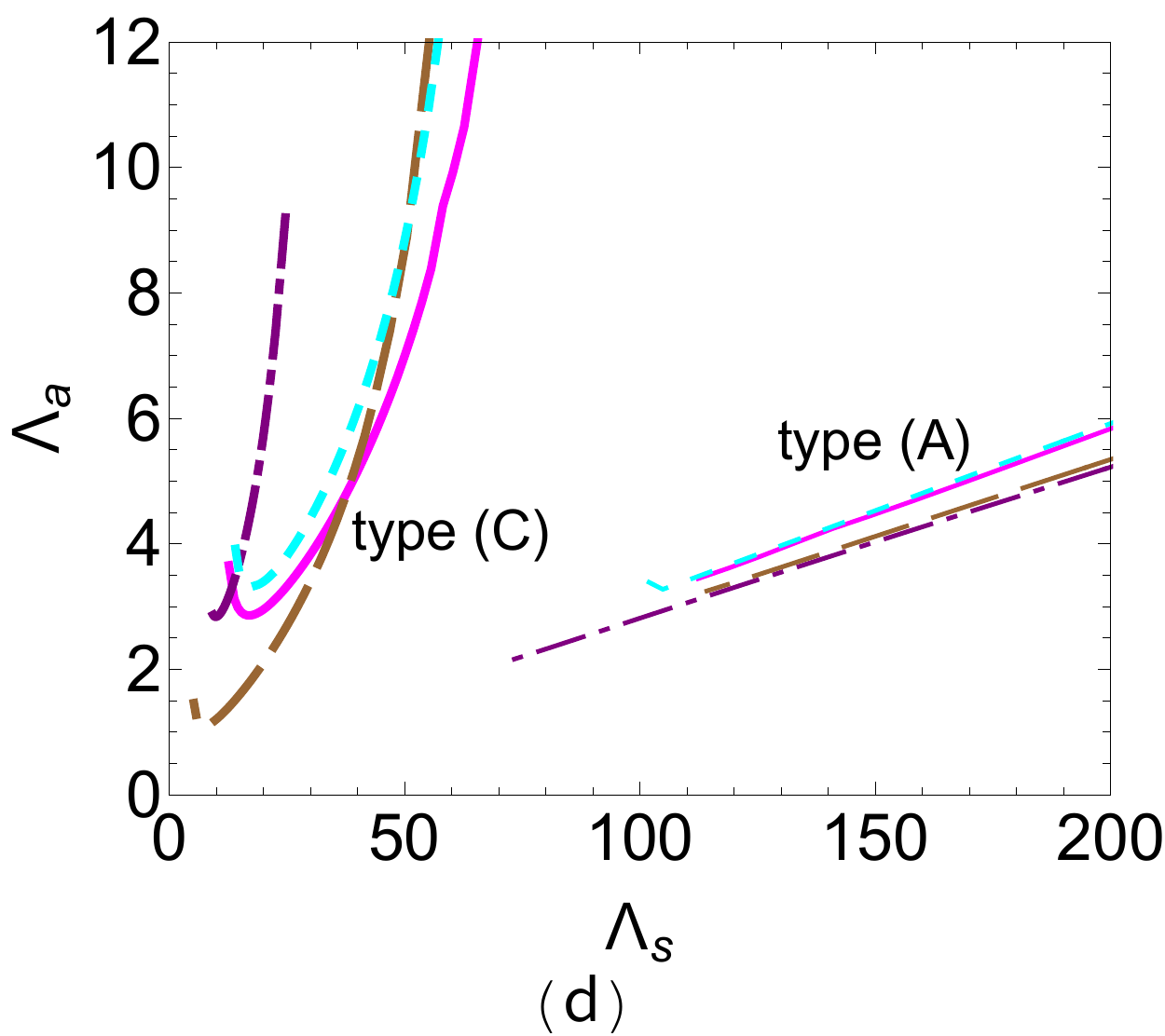}
\caption{Upper panels: Speed of sound (left) and mass-radius diagram (right) for for different equations of state. First-order phase transitions ($c_S = 0$) introduce a second stable branch in the mass-radius curves. Lower panels: BLRs between stars (with a mass ratio $0.75$) in the same branch (types A and C) or in different branches (types B.1 and B.2) produce a slope, hill, drop, and swoosh. Figure modified from Fig.1 in~\cite{Tan_20222}.} \label{FIG5}
\end{figure}

Another observational test that may be used to constrain equations of state is the evaluation of tidal deformabilities in neutron stars inferred from gravitational-wave measurements. In a coalescing binary of neutron stars, the gravitational field of one star perturbs the field of the other (and vice-versa), causing an acceleration in their inspiral. This change in the inspiral rate shapes the gravitational-wave emitted, and this wave thus provides information about the tidal deformabilities $\Lambda_{1,2}$ of the neutron stars. Considering a sequence of central densities for a given EoS and a fixed mass ratio, one can construct the \textit{binary Love relations} (BLRs) $\Lambda_s$ and $\Lambda_a$, definined with the symmetric and anti-symmetric tidal deformabilities $\Lambda_{s,a} = (\Lambda_1 \pm \Lambda_2)/2$. Due to phase transitions and the consequent non-smooth structure of the speed of sound $c_S$, which may tilt the mass-radius diagram, peculiar structures (such as slopes, hills, drops and swooshes) are created in the BLRs~\citep{Tan_20222}, as shown in Figure \ref{FIG5}. 

The change in slope in the BLRs may be observable already during the fifth LIGO observing run if a sufficiently loud and low mass neutron-star binary is detected. The detection of drops and swooshes is more challenging, because both occur at very small $\Lambda_a$, and such detection would require very low uncertainties in the measurements, which are achievable only if an exceptionally loud signal is detected.  

\section{Conclusions}

From the recent developments here reported, one my infer that new tight constraints from experiment, observation and theory
are slowly determining dense matter and neutron-star core
properties. In this context, EoS repositories (such as CompOSE and MUSES) help speeding up the understanding of dense matter. Furthermore, astrophysical constraints must be taken into account and, in this context, gravitational waves are providing new ways to study the dense matter EoS. Besides the basic mass-radius relation of neutron stars, more specific and subtle quantities (such as the speed of sound and tidal deformabilities for these objects) can be used to probe different equations of state. Advances in this field can be expected shortly, since LIGO, Virgo, and KAGRA are coordinating a new observing run in March 2023. Thus, open questions in nuclear astrophysics may soon find their answers and induce further interrogations about the intimate structure of matter.  

\section*{ACKNOWLEDGEMENTS}
V. D. acknowledges support from the National Science Foundation under grants PHY1748621, MUSES OAC-2103680, and NP3M PHY-2116686. R.L.S.F. acknowledges support from Conselho Nacional de Desenvolvimento Cient\'ifico 
e Tecno\-l\'o\-gico (CNPq), Grant No. 309598/2020-6 and Funda\c{c}\~ao de Amparo \`a Pesquisa do Estado do Rio Grande do Sul (FAPERGS), Grants Nos. 19/2551- 0000690-0 and 19/2551-0001948-3.

\bibliography{Dexheimer}

\end{document}